\documentclass{kluwer}    
\usepackage{psfig}
\usepackage{graphicx}

\newdisplay{guess}{Conjecture}

\begin{document}                                                                                   
\begin{article}
\begin{opening}         
\title{Search for and study of extremely  metal-deficient galaxies}
\author{S.A. \surname{Pustilnik}\thanks{Partially supported by SOC}\email{sap@sao.ru}}
\institute{Special Astrophysical Observatory of RAS, Nizhnij Arkhyz, 369167, Russia}
\author{A.Y. \surname{Kniazev}\email{kniazev@mpia.de}}
\institute{Max Planck Institut f\"{u}r Astronomie, K\"{o}nigstuhl 17, D-69117, Heidelberg, Germany}
\author{A.G. \surname{Pramskij}\email{pramsky@sao.ru}}
\author{A.V. \surname{Ugryumov}\email{and@sao.ru}}
\institute{Special Astrophysical Observatory of RAS, Nizhnij Arkhyz, 369167, Russia}
\runningauthor{S.A. Pustilnik et al.}
\runningtitle{Search for and study of extremely  metal-deficient galaxies}
\date{}

\begin{abstract}
 We summarize the progress in identifying and observational study of
  extremely metal-deficient (XMD) gas-rich galaxies (BCGs, dIr and LSBDs).
  Due to volume limitation only following issues are addressed: sample
  creation, some statistical data, Colour-Magnitude Diagrams (CMD) and ages,
  the case of
  SBS~0335--052 system, and probable evolution paths of various XMD objects.
\end{abstract}
\keywords{galaxy evolution}

\end{opening}           

\vspace{-0.5cm}

\section{Introduction}

XMD gas-rich galaxies (conditionally with 12+$\log$(O/H) $<$ 7.65),
are very rare objects in Local Universe, which best approximate the
properties of young galaxies formed $\sim$13 Gyr ago. Since real young
galaxies at that epoch are too faint to study them in detail, their local
cousins provide a valuable information on the complex process of galaxy
formation and early evolution. Moreover, some of known XMD gas-rich galaxies
can be truly young local galaxies, just recently past the phase of
protogalaxy. The specifics of XMD galaxy properties can be understood from
comparison with the properties of more typical galaxy samples. Therefore the
parallel study of main properties of general BCG, LSB and dI samples is
very important.

\vspace{-0.5cm}
\section{XMD galaxy sample creation}

In the beginning of 90-ties we realized that to understand the nature of XMD
galaxies as a group we need to study the properties of sufficiently large
number of such galaxies with reliable estimates of O/H. Besides of well known
Second Byurakan Survey, the search for new XMD BCGs was in particular the
goal of Hamburg-SAO survey (HSS) for ELGs (Ugryumov et al. 2001 and references
therein), and HSS-LM (Ugryumov et al. 2002).
Kunth \& \"Ostlin (2000) gave a compilation of XMD galaxies  known in 1999,
including 18 BCGs, 6 LSBDs and 7 dIs. dIs are from the Local Group
(LG) and its environments. Since that time 4 BCGs of their list were
revisited and rejected. 16 more BCGs are added (many of which are still in
press or preparation), thus increasing the number of XMD BCGs till 30. 9 of
them are from the HSS for ELGs and HSS-LM. 4 new XMD BCGs are
from KISS (Melbourne \& Salzer 2002). Only one LSBD (UGCA 292) was added
(van Zee, 2000).

\section{Some Statistical Data}

Surface density of XMD BCGs, as revealed by systematical ELG surveys
(combined SBS+HSS data) is $\sim$4 per 1000~sq.deg.
(for $B \leq 18.5^m$, 12+$\log$(O/H) $\leq$ 7.63).
Over the whole unobscured sky ($|b^{II}|$ $\le 20^{\circ}$) $\sim$120 XMD BCGs
is expected till $B$$\sim$18.5$^m$.
Six new XMD BCGs are found in {\it Voids} and similar environmnets. Is this
a hint on special conditions for very slow or retarded formation/evolution ?

\begin{figure*}
\centering
\includegraphics[width=4.0cm,angle=-90]{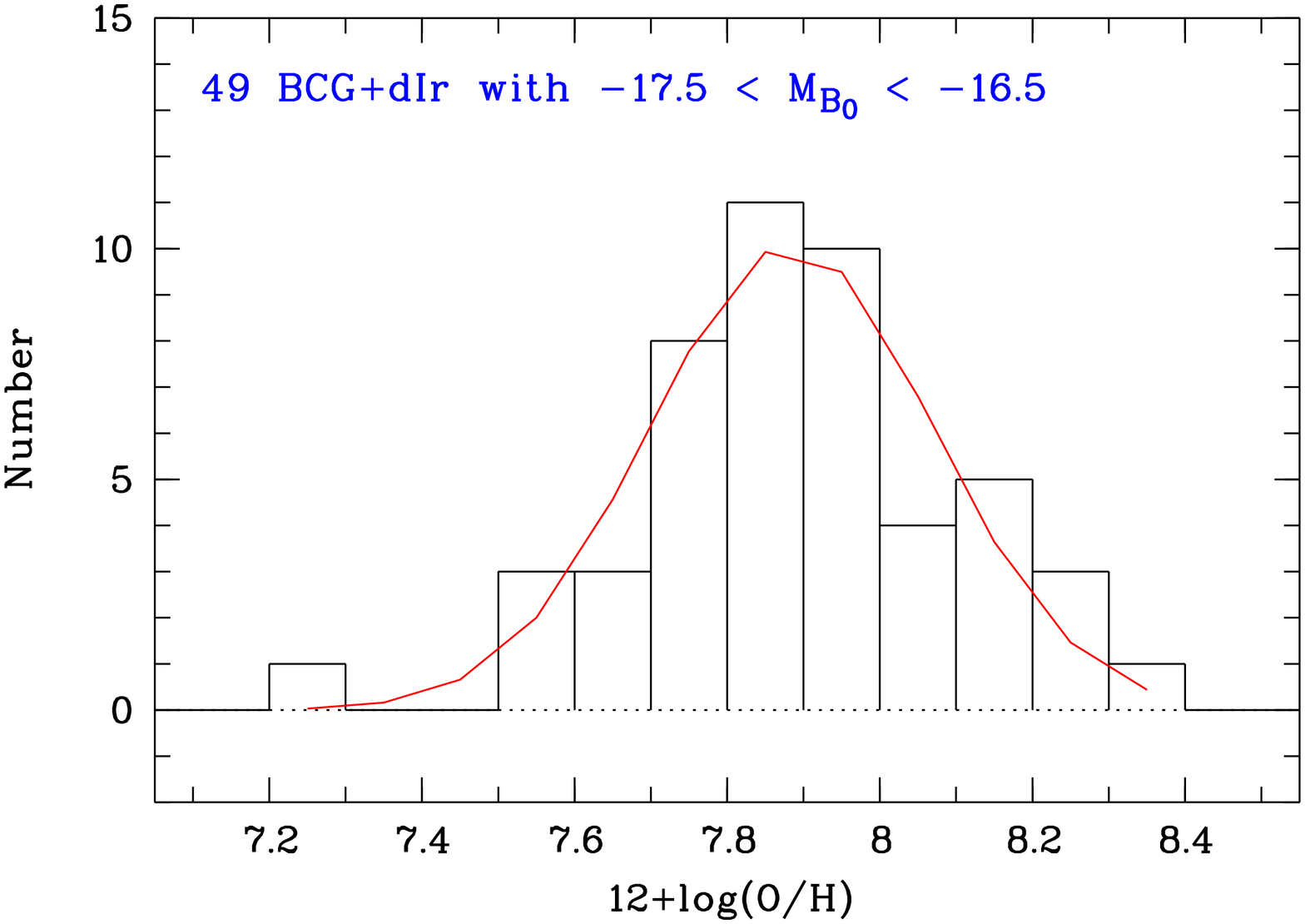}
\includegraphics[width=4.0cm,angle=-90]{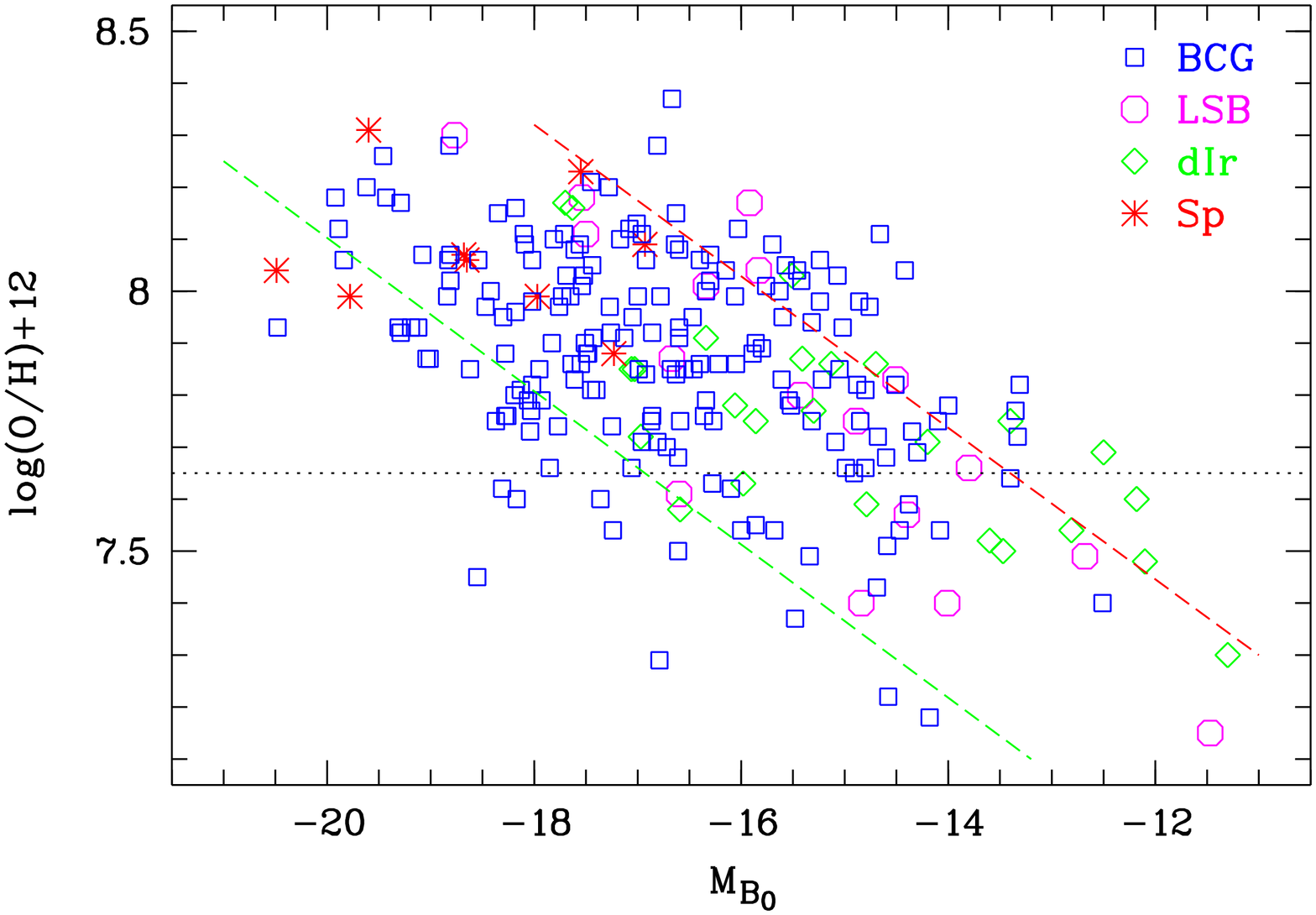}
\caption{\underline{\it Left panel}: Histogram of O/H distribution for 49
BCGs and dIs with $M_\mathrm{B}$ near --17$^{m}$. Solid line shows a Gaussian
fit. The outlier is SBS~0335--052~E.
\underline{\it Right panel}: O/H vs $M_{B}$ for 236 galaxies with
$\sigma$(O/H) $\le0.10$ dex.
}
 \label{Fig1}
\end{figure*}

Distribution of O/H can be a useful indicator of the nature of XMD BCGs. If
all observed galaxies are from the same ensemble started star formation
$\sim$13 Gyr ago with different SF rates and metal loss efficiencies, one
should expect smooth unimodal O/H distribution with a tail on small O/H.
If small, but significant fraction of truly young XMD galaxies exists, this
should result in a second peak at very low O/H, and hence in bimodality of
O/H distribution. While the existing statistics is quite limited, there are
hints on such bimodality (see, e.g., left panel of Fig.~\ref{Fig1}).
Similar hint is present in plot of O/H vs M$_\mathrm{B}^{0}$, drawn on
236 galaxies (right panel of Fig.~\ref{Fig1}, BCGs are
mainly from SBS, HSS-ELG and HSS-LM).
While the well known trend is still noticeable, the scattering at very
low O/H reaches 6 magnitudes, suggesting non-typical evolution histories
for the most luminous XMD galaxies.

\section{Resolved stars and old XMD galaxies}

SF histories of the LG XMD dIr were studied with 
CMDs of resolved stars. All of them have detectable old populations. Due to
lack of space we cite only review by Mateo (1998). LG XMD galaxies are, thus,
certainly old. New HST data evidence for old stars
in a bit more distant XMD LSBDs: DDO 53 (L.Makarova, priv. communication)
and UGCA 292 (van Zee, priv. communication). Izotov \& Thuan (2002) for
UGC~4483 found stars with  $T\sim$2~Gyr.
The most metal-deficient BCG I~Zw~18 ($Z \sim$1/51~Z$_{\odot}$) was claimed
to have old stars (\"Ostlin 2000 and references therein) on HST images.
However this issue did not settled yet due to the problems with the
proper account of clumpy dust extinction.

\begin{figure*}
\centering
\includegraphics[angle=0,width=12.0cm,bb=29 458 560 714,clip=]{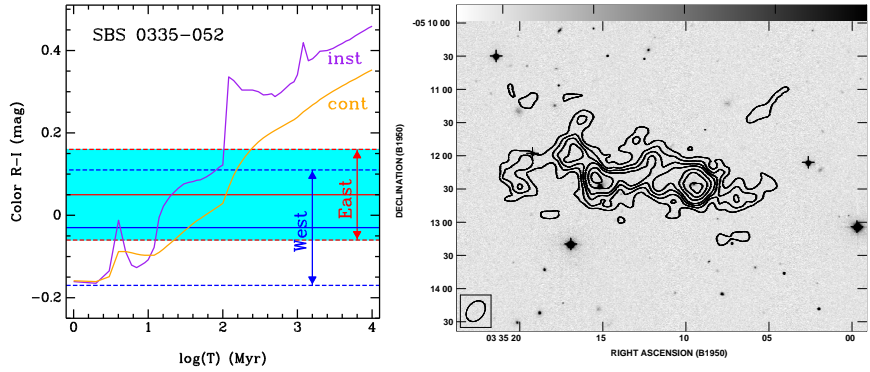}
\caption{\underline{\it Left panel}: $(R-I)$ colours vs age for PEGASE.2
evolution tracks (Fioc \& Rocca-Volmerange 1997)
with superimosed $(R-I)$ colours of underlying emission of
E and W components of SBS~0335--052, after the gas emission subtraction.
Arrows indicate the range $\pm1\sigma$ for derived $(R-I)$.
\underline{\it Right panel}: VLA 21-cm map of giant disturbed H{\sc i} cloud,
in which both Eastern and Western galaxies are immersed.
}
\label{Fig2}
\end{figure*}

\section{The case of XMD pair SBS~0335--052~E and W}

To estimate the ages of unresolved stars in the outermost regions of XMD BCGs
the subtraction of ionized gas emission is often crucial. The SAO 6\,m
telesope deep  $UBVRI$,H$\alpha$ surface photometry of SBS~0335--052~E,W
(Pustilnik et al. 2002, in prep.) results in {\bf no} traces of stellar
populations with ages T$>$200 Myr.
Comparison of derived gas-emission subtracted colours with the PEGASE.2
evolution models tracks is shown in the left panel of Fig.~\ref{Fig2}.
The similar  analysis of I~Zw~18 by Papaderos et al. (2002)
results in the similar conclusions.

The W component of the system SBS~0335--052 (at $\sim$22 kpc from E component)
is a dwarf galaxy with O/H close to that of I~Zw~18 (Lipovetsky et al. 1999).
Its pairing with the E galaxy favors a hypothesis of their
common recent formation in one huge H{\sc i} cloud disturbed
by an external galaxy (see VLA H{\sc i} map from Pustilnik et al. 2001 in
Fig.~\ref{Fig2}, and discussion therein).
A chance collision of two such unusual dwarfs looks highly unprobable due
to their very low space density.


\section{Evolution paths of XMD galaxies}

While the amount of observational data on XMD galaxies is far from
sufficient for firm conclusions on their evolution status/scenario,
some preliminary options emerge from the data accumulated to-date:
\begin{itemize}
\item Extremely gas-rich LSB galaxies outside the Local Group -- DDO~154
  (Kennicutt \& Skillman 2001) and UGCA~292 (van Zee 2000), evolve very
  slowly well along the closed-box model track.
\item Local Group  XMD dIr galaxies have either the deficit of metals for
     their gas content (due to galactic winds), or stripped gas for their low
     metallicity (Kennicutt \& Skillman 2001). They all are old objects.
\item There are a few XMD BCGs with very blue LSB hosts. Probably some of
      them are young (e.g., SBS~0335--052~E and W, I~Zw~18).
\end{itemize}

\vspace{-0.5cm}
\acknowledgements
S.A. Pustilnik is grateful to the organizers for invitation and
partial financial support to attend the conference, and to J.Salzer for
sending some data prior publication.

\end{article}
\end{document}